\documentclass{article}

\usepackage{amsmath,amssymb,amsfonts,epsfig,psfrag,latexsym,lscape}
\usepackage{booktabs}
\usepackage[usenames,dvipsnames]{color}
\usepackage{cite,hyperref}

\usepackage[top=3cm, bottom=2.5cm, left=2.5cm, right=2.5cm]{geometry}

\newcommand{\la}{\langle}
\newcommand{\ra}{\rangle}

\newcommand{\cO}{{\cal O}}
\newcommand{\cA}{{\cal A}}
\newcommand{\cB}{{\cal B}}

\newcommand{\cN}{{\cal N}}

\newcommand{\tr}{\mathrm{Tr}}

\begin{document}

\thispagestyle{empty}

\begin{flushright}
HU-EP-16/25 
%, HU-MATH 2016-14
\end{flushright}

\begin{center}

\vskip 4.5 cm

{\Large \bf Tessellating cushions: four-point functions in $\cN=4$ SYM}

\vskip 2 cm
 
{\bf Burkhard Eden${\,}^1$, Alessandro Sfondrini${\,}^2$} \\

\vskip 1 cm

{\it ${\,}^1$ Institut f\"{u}r Mathematik und Physik, Humboldt-Universit{\"a}t zu Berlin, \\
Zum gro{\ss}en Windkanal 6, 12489 Berlin, Germany;\\[0.2cm]
${\,}^2$ Institut f\"ur Theoretische Physik, ETH Z\"urich,\\
Wolfgang-Pauli-Str.~27, 8093 Z\"urich, Switzerland.}

\vskip 1 cm

E-mails: eden@math.hu-berlin.de, sfondria@itp.phys.ethz.ch.

\end{center}

\vskip 2.5 cm
\noindent
We consider a class of planar tree-level four-point functions in $\cN = 4$ SYM in a special kinematic regime: one BMN operator with two scalar excitations and three half-BPS operators are put onto a line in configuration space; additionally, for the half-BPS operators a co-moving frame is chosen in flavour space.
In configuration space, the four-punctured sphere is naturally triangulated by tree-level planar diagrams. We demonstrate on a number of examples that each tile can be associated with a modified hexagon form-factor in such a  way as to efficiently reproduce the tree-level four-point function.
Our tessellation is not of the OPE type, fostering the hope of finding an independent, integrability-based approach to the computation of planar four-point functions.

\newpage

\section{Introduction}
It is well-known that the spectrum of single-trace operators in  $\cN = 4$ super-symmetric Yang-Mills (SYM) can be described by an integrable system, see~\cite{Beisert:2010jr} for a review. A more recent development is the observation that non-extremal three-point functions of single-trace operators can also be studied using integrability. This was first worked out in terms of a spin-chain ``tailoring'' picture at tree-level~\cite{Escobedo:2010xs} and then extended to all loops by the ``hexagon'' approach~\cite{Basso:2015zoa}. While the hexagon prescription is not yet complete---finite-size corrections akin to L\"uscher corrections in integrable field theories~\cite{Luscher:1985dn,Luscher:1986pf} need to be more systematically understood~\cite{Eden:2015ija,Basso:2015eqa}---its success for a large number of correlators strongly suggests that the integrability structure can be used to compute three- and possibly higher-point functions.

A natural next step are four-point functions. Unlike two- and three-point functions, whose space-time dependence is fixed by conformal invariance, these depend on the position of the operators through two \textit{conformal cross-ratios}. In fact, this dependence is highly non-trivial: at every loop order there will be a new polylogarithmic function of the cross-ratios, even in the simplest case of a four-point function of half-BPS operators (whose two- and three-point functions are entirely protected from quantum corrections).
Four-point functions are in principle uniquely determined in terms of the conformal data through the operator-product expansion (OPE). One might indeed think of combining the integrability approach with the OPE to construct four-point functions. While this is certainly worth investigating, the difficulties in re-summing the whole OPE make it desirable to look for different approaches. The aim of this letter is to propose such an approach, exploiting the hexagon formalism.

We will restrict to a special configuration. Consider $K$ scalar single-trace operators on a line. If the operators are half-BPS \textit{and suitably rotated in R-symmetry space}, the resulting $K$-point function is protected, as argued by Drukker and Plefka~\cite{Drukker:2009sf}. For $K=3$, this  configuration is indeed the  vacuum of the hexagon approach; we will also take it as a starting point here, even if for $K=4$ such a set-up is non-generic. We then consider the case in which one of the half-BPS operators is replaced by a non-protected Berenstein-Maldacena-Nastase (BMN) operator~\cite{Berenstein:2002jq}. Then we expect the four-point function to depend on the position as well as on the structure constants of the operators involved; this dependence is what we will compute at tree-level for single-trace operators.

\begin{figure}[t]
\includegraphics[width = \linewidth]{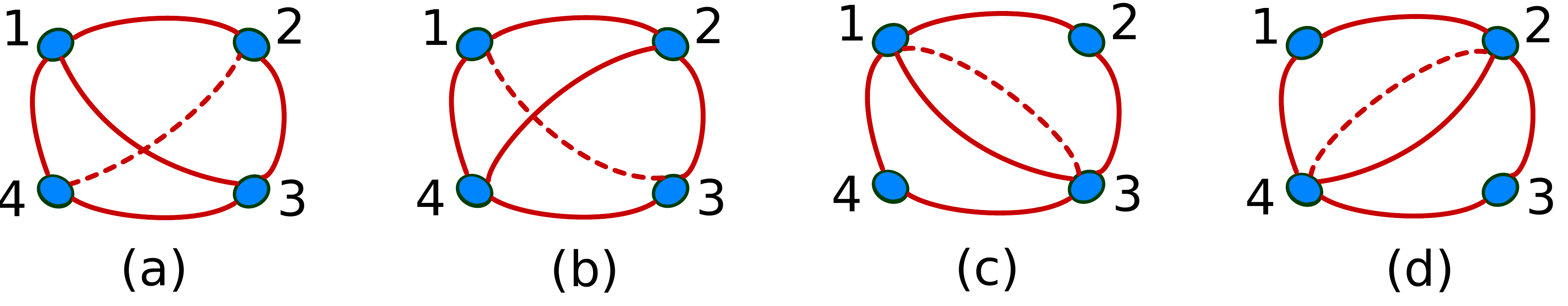}
\caption{Planar four-point tree-level diagrams involving single-trace operators. Operators are depicted by a black ring, with crimson lines between operators $i,j$ representing the product of $l_{ij}$ propagators.}
\label{fig:topologies}
\end{figure}

Our proposal is inspired by integrability and  guided by the structure of planar tree-level diagrams in $SU(N)$ gauge theory: the fields in a single-trace operator are like ``beads on a chain'', to be connected to those of the other operators by free propagators on the surface of a sphere. Line-crossings are suppressed in the planar limit. As a consequence, for the four-point function one can draw only four types of graphs, each of which provides a tiling of the sphere by four triangles, cf.~Figure~\ref{fig:topologies}. We can think of the operators as sitting at the corners of a soft cushion with propagators connecting them along the edges. The diagrams acquire volume by marking a diagonal on the front (solid line) and another one on the back of the cushion (dashed lines). Planarity does not permit any other configuration. 

We want to compute these four-point functions by partitioning each of the topologies of Figure~\ref{fig:topologies} into four hexagons. Each hexagon will have three edges running in parallel to the propagators (crimson  lines) and three edges on a portion of the chain representing the operators (black circles). We can then use a suitable modification of the formalism introduced by Basso, Komatsu and Vieira (BKV)~\cite{Basso:2015zoa} to compute the hexagon form factors\footnote{For this purpose it is important to take all our operators on a line, which guarantees that each set of three operators lies on a line too, as it is necessary in the hexagon approach.}.  While the original approach cleanly separates three-point functions into a space-time factor and the structure constant---computing the latter though hexagon form-factors---we rather put a part of the space time factor into the hexagon vertices in order to reproduce the (restricted) kinematic dependence of our mixed BMN-BPS${^3}$ correlators. We then proceed to classify all possible tree-level diagrams of topology (a) to (d) (cf.~Figure~\ref{fig:topologies}) by conformal weights, and evaluate their overall numerical factor and kinematic dependence by integrability.

In this way, we reproduce a number of tree-level four point functions of single-trace BMN-BPS${^3}$ operators; we checked our proposal for two-impurity BMN operators of length up to seven. It is worth stressing that, already at tree-level, this approach appears quite efficient compared to straightforward Wick contractions. Our method is similar in spirit to the one proposed by Caetano and Escobedo \cite{Escobedo:2010xs}, which relied on computing spin-chain scalar products---the perturbative ``tailoring'' picture. The fact that we instead reduce  the four-point functions to a hexagon tessellation gives us confidence that our approach may be extended to more general operators and higher loop orders---as the hexagon form factor is known at all loops and for generic operators.

%The spectrum problem in $\cN = 4$ SYM has been described by an integrable system \cite{MZ,beisStau} a  while ago. Current work in the domain is directed at higher-point objects in the model. The so-called remainder function in scattering amplitudes can by now also be addressed \cite{WLOPE,...} and very recently the ``hexagon proposal'' for three-point functions of gauge invariant composite operators has been formulated \cite{BKV}.

\section{BMN operators and tree-level four point functions}

\subsection{BMN operators at one loop}
Here we focus on a class of single-trace ``BMN operators'' \cite{Berenstein:2002jq} that yield a simple spin-chain picture in planar $\cN = 4$ SYM. Let $Z$ a complex scalar; then
\begin{equation}
\cO_L = \frac{1}{\sqrt{L N^L}} \tr(Z^L).
\end{equation}
These scalar operators are half-BPS and highest-weight states of the $su(4)$ R-symmetry algebra; their dimension does not receive quantum corrections. An immediate generalization of these operators is to insert two impurities%
\footnote{Inserting a single impurity results in a R-symmetry descendant of the highest-weight state, due to the cyclicity of the trace.}
\begin{equation}
\cO_L^k = \frac{1}{\sqrt{N^L}}\tr(Z^{L-k-2} Y Z^k Y),
\end{equation}
where $Y \neq \bar Z$ is another scalar field; this probes an $su(2)\subset su(4)$ sector of the theory. 
Due to the cyclicity of the $SU(N)$ colour trace there are $[L/2]$ distinct operators at length $L$. Already at the first order of perturbation theory, these acquire a non-trivial anomalous dimension encoded in a mixing matrix. It is well known that this mixing problem is equivalent to the diagonalization of the Heisenberg spin-chain Hamiltonian~\cite{Minahan:2002ve}. The spectrum is constructed out of magnons, or spin-waves; these two-impurity states depend on two momenta $p_1,p_2$  with associated ``rapidities''  $u_j = \frac{1}{2} \cot\left(\frac{p_j}{2}\right)$. Integrability of the Heisenberg Hamiltonian imposes the Bethe Ansatz equations $
e^{ip_jL}S_{kj}=1$ with $j\neq k=1,2$
and
\begin{equation}
\label{eq:smat}
e^{ip_j}=  \frac{u + \frac{i}{2}}{u - \frac{i}{2}},
\qquad
S_{jk}%\equiv S\big(p(u_j),p(u_k)\big)
=\frac{u_j - u_k + i}{u_j - u_k - i}.
\end{equation}\\
Furthermore, cyclicity imposes $p_1+p_2=u_1+u_2=0$, yielding that physical states are determined by the quantization condition
\begin{equation}
\left( \frac{u + \frac{i}{2}}{u - \frac{i}{2}} \right)^{L-1} = 1,
\end{equation}
with $u\equiv u_1=-u_2$. The one-loop anomalous dimension is the spin-chain energy level
\begin{equation}
\gamma_1 =  E(u_1)+E(u_2)=2 E(u)=\frac{2}{u^2 + \frac{1}{4}}.
\end{equation}
We collect the first few eigenstates of the one-loop dilatation operator in Table~1, along with their anomalous dimension and rapidity~$u$.%
\footnote{More precisely, we write the single-trace part of the operators. $\cB_{4,5}$ receive no double trace admixtures while at length 6 and 7 there is mixing with $\cO_l\cdot \cB_{4,5}$.  Here we also ignore the descendants of the BPS operators, whose anomalous dimension trivially vanishes.}

\begin{table}[t]
\begin{center}
\begin{tabular}{l|l|l|c}
$L$ & Conformal eigenstate & $\gamma_1$ & $u_1$ \\[1 mm]
\midrule
4 & $\cB_4= \frac{1}{\sqrt{3}}\Big[\cO_4^0 - \cO_4^1\Big]$ & 6 & $\frac{1}{2 \sqrt{3}}$ \\[2 mm]
5 & $\cB_5=\frac{1}{\sqrt{2}} \Big[\cO_5^0 - \cO_5^1\Big]$ & 4 & $\frac{1}{2}$ \\[2 mm]
6 & $\cB_6^\mp=\frac{1}{\sqrt{5}}\Big[\frac{1}{2} (1 \pm \sqrt{5}) \cO_6^0 + \frac{1}{2} (1 \mp \sqrt{5}) \cO_6^1 - \cO_6^2\Big]$ & $5 \mp \sqrt{5}$ & $\frac{1}{2} \sqrt{ 1 \pm \frac{2}{\sqrt{5}}}$ \\[2 mm]
7 & $\cB_7'=\frac{1}{\sqrt{2}}\Big[\cO_7^0 - \cO_7^2\Big]$ & 2 & $\frac{\sqrt{3}}{2}$ \\[2 mm]
7& $\cB_7''=\frac{1}{\sqrt{6}}\Big[\cO_7^0 - 2 \cO_7^1 + \cO_7^2\Big]$ & 6 & $\frac{1}{2 \sqrt{3}}$
\end{tabular}
\caption{We list the two-impurity BMN operators that acquire a non-trivial anomalous dimension with length from 4 to 7. We denote the operators by $\cB_L^k$, and indicate their anomalous dimension $\gamma_1$ and the related Bethe Ansatz rapidity $u$.}
\end{center}
\end{table}

\subsection{Four-point functions}
A very simple (and in fact, protected) set of four-point functions  was considered in Ref.~\cite{Drukker:2009sf}. We take four half-BPS operators on a line parametrized by $a\in\mathbb{R}$, and perform an $a$-dependent $su(4)$ rotation in R-symmetry space of the form
\begin{equation}
\hat{Z}(a) = \big[Z + a^2 \bar Z + a (Y - \bar Y)\big] \, , \label{rotate}
\end{equation}
or, in an $so(6)$ covariant notation 
\begin{equation}
\hat Z(a) = z^\mu_a \phi_\mu \, , \quad z^\mu_a = \left( (1 + a^2), 0, 0, 2 a, 0, i (1 - a^2) \right) \, , \label{rotate2}
\end{equation}
viewing the fifth and, after Wick rotation, the sixth component as times as in the embedding formalism; the middle four entries parametrise Minkowski space. Labelling the four points on a line by $a_i$ with
$x_i^\nu = \delta^\nu_3 \, a_i$ we find
\begin{equation}
\la \hat Z(a_1) \hat Z(a_2) \ra = \frac{(a_1 - a_2)^2}{4 \pi^2 (a_1 - a_2)^2} = \frac{1}{4 \pi^2} \, . \label{prop}
\end{equation}
In the following we will drop the factor $4 \pi^2$.

We now want to consider a non-protected four-point function. To this end, we replace one of the half-BPS operators with a BMN operators of the type $\cB_L^k$. Without loss of generality, let us place it at $a_1 = 0$.
Note that $\la Z(0) \hat Z(a_2) \ra = 1$ and $\la Y(0) \hat Z(a_2) \ra = - \frac{1}{a_2}$; hence the four-point functions we wish to construct will be homogeneous of order $-2$ in the $a_i$. More is true: we can rewrite all our free field theory results in terms of two differences, say
\begin{equation}
a_{23} = \frac{1}{a_2} - \frac{1}{a_3} \, , \qquad a_{34} = \frac{1}{a_3} - \frac{1}{a_4} \, .
\end{equation}
By way of example we introduce the notation
\begin{equation}
G(7'';2,3,2) = \la \cB_7''(0) \cO_2(a_2) \cO_3(a_3) \cO_2(a_4) \ra \, .
\end{equation}
Wick contractions yield the results in Table 2.
\begin{table}[ht]
\begin{center}
\begin{tabular}{l|cc}
& $C$ & \underline{v} \\
\midrule
$G(4;222)$ & $4 \sqrt{\frac{2}{3}} $&$ (1, 1, 1)$ \\[2 mm]
$G(4;242)$ & $\frac{8}{\sqrt{3}} $&$ (1, 0, 1)$ \\[2 mm]
$G(4;233)$ & $\sqrt{6} $&$ (2, 2, 3)$ \\[2 mm]
$G(5;232)$ & $\sqrt{6} $&$ (3, 2, 3)$ \\[2 mm]
$G(6^\mp;222)$ & $4 \sqrt{2} $&$ (1, 1, 1)$ \\[2 mm]
$G(4;235)$ & $\sqrt{10} $&$ (2, 4, 5)$ \\[2 mm]
$G(4;244)$ & $8 \sqrt{\frac{2}{3}} $&$ (1, 1, 2)$ \\[2 mm]
$G(4;343)$ & $2 \sqrt{3} $&$ (3, 2, 3)$ \\[2 mm]
$G(5;252)$ & $3 \sqrt{10} $&$ (1, 0, 1)$ \\[2 mm]
$G(5;234)$ & $2 \sqrt{3} $&$ (3, 4, 7)$ \\[2 mm]
$G(5;333)$ & $9 \sqrt{6} $&$ (1, 1, 1)$ \\[2 mm]
$G(6^\mp;242)$ & $\frac{4 (1 \pm \sqrt{5})}{\sqrt{5}} $&$ (2, 1, 2)$ \\[2 mm]
$G(6^\mp;233)$ & $\frac{3 (1 \pm \sqrt{5})}{\sqrt{10}} $&$ (4, 4, 6 \pm \sqrt{5})$ \\[2 mm]
$G(7';232)$ & $2 \sqrt{6} $&$ (2, 1, 2)$ \\[2 mm]
$G(7'';232)$ & $6 \sqrt{2} $&$ (1, 1, 1)$
\end{tabular}
\caption{Tree-level four-point functions of single-trace operators, computed by Wick contraction are encoded in a prefactor $C$ and a vector $\underline{v}$; namely $G(\cdots) = C\, \underline{v}\cdot \underline{a}$ where the position vector is $\underline{a}=(a_{23}^2, \, a_{23} a_{34}, \, a_{34}^2)$.}
\end{center}
\end{table}
We have evaluated correlators that require up to seven Wick contractions.\footnote{We have restricted to planar contractions of single-trace operators. This is consistent at large-$N$ with the possible exception of the correlator $G(7'';232)$. It is interesting that we can reproduce the planar and single-trace contribution to this operator by the hexagon method, suggesting that such an approach works across the single-trace sector.}
This can, of course, be improved a little without too much difficulty, although the fact that every rotated vacuum $\hat Z$ contains four terms renders the enterprise somewhat clumsy already at this level. Note that $G(4;262) = 0$ for group theory reasons. The leading order four-point function at large $N$ goes like $1/N^2$; we omit this overall factor.
%Further, for $n$ Wick contractions, the connected part\footnote{There cannot be any disconnected pieces because exactly one of the operators belongs to a long multiplet.} of the planar correlators is of order $1/N^2$. We suppressed this overall factor, too.

The pattern of numbers in the examples we studied is far from easy to understand. In the next section we want to derive it by integrability methods. The aim is to list all possible planar tree-level graphs on the sphere for operators of lenghts $L_1,L_2,L_3,L_4$, and to associate coefficients and kinematic dependence to each of the diagrams using the recently developed hexagon operators.

\section{The hexagon approach and four-point functions}

\begin{figure}[t]
\begin{center}
\includegraphics[width = 0.3\linewidth]{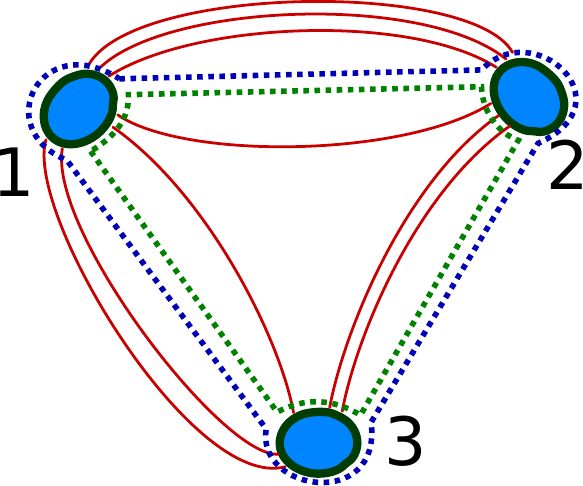}
\end{center}
\caption{A planar three-point function can be split into an inner and an outer hexagon (green and blue dotted lines, respectively). This splitting partitions the Wick contractions (crimson solid lines) among the two hexagons. In the BKV proposal, we should sum over all these partitions with appropriate weights.}
\label{fig:hexagon}
\end{figure}
\subsection{The hexagon approach}
Planar three-point functions in $\cN=4$ SYM can be computed by the hexagon approach~\cite{Basso:2015zoa}. Let us represent a three-point function as a triangle; three spin-chain states are the vertices and the propagators are the edges, as in figure~\ref{fig:hexagon}. Then cutting this figure in two yields two hexagons, with three edges made out of pieces of the spin-chain, and three ``virtual'' edges along the propagators. The tree level three-point function can be obtained by summing over all the possible ways of distributing the spin chain excitations over the hexagons---this precisely reproduces the ``tailoring'' picture of Escobedo, Gromov, Sever and Vieira~\cite{Escobedo:2010xs}. A key advantage of the ``hexagon'' picture over the ``tailoring'' approach is that the former can be related to a form-factor on the worldsheed of the dual AdS$_5\times$S$^5$ string theory. This allows us to use the integrability of the world-sheet S~matrix to obtain an all-loop prediction for the three-point function of operators with large R-charge.%
\footnote{The hexagon picture is obtained in an asymptotic regime, where finite-size corrections due to virtual particles wrapping cycles in the worldsheed can be overlooked. It is an ongoing struggle to incorporate such virtual-particle effects by L\"uscher-like corrections~\cite{Eden:2015ija,Basso:2015eqa}.
} Let us review this approach in the simple case where only one of the three operators is non-protected, and contains excitations in an $su(2)$ sector only---this will be sufficient for our four-point function construction. We denote the set of all magnon rapidities as ${\bf u}$; for us, ${\bf u}=\{u_1,u_2\}$. Then we express the three-point function $\langle \cB_{L_1}^{k}\cO_{L_2}\cO_{L_2}\rangle$ in terms of the one of three half-BPS operators $\langle \cO_{L_1}\cO_{L_2}\cO_{L_2}\rangle$,
\begin{equation}
\label{eq:BKVformula}
\frac{\langle \cB_{L_1}^{k}\cO_{L_2}\cO_{L_2}\rangle}{\langle \cO_{L_1}\cO_{L_2}\cO_{L_2}\rangle}=\frac{\mathcal{A}}{\sqrt{\mathcal{G}}\prod_{j<k}\sqrt{S_{jk}}},
\end{equation}
with 
\begin{equation}
\mathcal{A}=\sum_{\bf{u}=\alpha\cup\bar{\alpha}} (-1)^{|\bar{\alpha}|}\, \omega(\alpha,\bar{\alpha},l_{12}) \,\mathfrak{h}_{\text{front}}(\alpha)\,\mathfrak{h}_{\text{back}}(\bar{\alpha}),
\end{equation}
where $\mathcal{G}$ is the Gaudin norm, $\mathfrak{h}$ is the hexagon form factor, $\mathfrak{h}(\alpha)=\prod_{j<k\in\alpha} h_{jk}S_{kj}$ and $\mathfrak{h}(\emptyset)=\mathfrak{h}(\{u_j\})=1$. At leading order, $S_{jk}$ is given by eq.~\eqref{eq:smat} and $h_{jk}=\frac{u_j-u_k}{u_j-u_k-i}$. The splitting-factor $\omega$ gives the relative normalization of the different partitions
\begin{equation}
\omega(\alpha,\bar{\alpha},l)=\prod_{k\in\bar{\alpha}} e^{ip_kl}\prod_{j\in\alpha, j>k}S_{kj},
\end{equation}
where $l$ is the so-called ``bridge-length'', i.e.~the number of tree-level contractions between a pair of operators in Figure~\ref{fig:hexagon}.

For our purposes it is convenient to slightly modify this Ansatz to make explicit the dependence on the coordinates $a_2,a_3$ (we have set $a_1=0$) by replacing $\mathcal{A}$ with $\widehat{\mathcal{A}}$,
\begin{equation}
\widehat{\mathcal{A}}=\sum_{\bf{u}=\alpha\cup\bar{\alpha}} \, \omega(\alpha,\bar{\alpha},l_{12}) \,\mathfrak{h}_{\text{f}}(\alpha)\,a_{23}^{|\alpha|}\;\mathfrak{h}_{\text{b}}(\bar{\alpha})\,a_{32}^{|\alpha|}\,.
\end{equation}
Note that this automatically accounts for the $(-1)^{|\bar{\alpha}|}$ sign. This prescription has a simple interpretation in terms of Wick contractions. As in figure~\ref{fig:hexagon}, the inner (``front'') hexagon connects points 1,2,3 (clockwise around the triangle), $\mathfrak{h}_{\text{f}}\approx\mathfrak{h}_{123}$; the outer one (``back'') connects $1,3,2$ (also clockwise) $\mathfrak{h}_{\text{b}}\approx\mathfrak{h}_{132}$. Excitations are only on operator~1, and can propagate to point 2 or 3. If $\mathfrak{h}(\alpha)$ is a hexagon with $|\alpha|$ excitations, then
\begin{equation}
\mathfrak{h}_{\text{f}}(\alpha) \to \mathfrak{h}_{\text{f}}(\alpha)\, a_{23}^{|\alpha|},
\qquad
\mathfrak{h}_{\text{b}}(\bar{\alpha}) \to \mathfrak{h}_{\text{b}}(\bar{\alpha})\, a_{32}^{|\bar{\alpha}|}.
\label{putKin}
\end{equation}
While this is inconsequential for the three-point function, it will be very useful to keep track of the dependence of each hexagon on the $a_j$ coordinates: as we will see, it will neatly yield the space-time dependence of the tree-level four-point function.
Finally, we should also bear in mind that the overall scaling of the half-BPS three point in eq.~\eqref{eq:BKVformula} function results in a factor of $\sqrt{L_1L_2L_3}$.

\subsection{Four-point function by hexagons}
We now want to compute tree-level four-point functions by tessellating each panel of figure~1 with four hexagons. We take operator~1 to be a BMN operator at $a_1=0$, while the other (half-BPS) operators sit at $a_j$.

Consider panel (d) in figure~1. The sphere is split into four hexagons, which we denote by their vertices, labelled clockwise. We therefore have hexagons $124$ (front), $142$ (back), $234$ (front) and $243$ (back). The latter two hexagons involve only half-BPS operators, therefore we expect the amplitude for this graph only to involve hexagons $124$, $142$; we define
\begin{equation}
\widehat \cA^{(2)} = \sum_{{\bf u} = \alpha \cup \bar \alpha} \omega(\alpha,\bar\alpha,l_{12}) \ \mathfrak{h}_{124}(\alpha) \, a_{24}^{|\alpha|} \ \mathfrak{h}_{142}(\bar\alpha) \, a_{42}^{|\bar \alpha|}.
\end{equation}
To avoid over-counting we should insist on $l_{24}^{\text{f}},l_{24}^{\text{b}} \geq 1$ for both edge-widths on the back and the front of the ``cushion''. Were this not the case, we would be able to represent this graph on the topology of panels (a) or (b) in figure 1, too. Note that with this restriction, graphs of type (d) only appear for correlators with 8 or more Wick contractions%
\footnote{ The shortest half-BPS operator has $L=2$ while the shortest BMN operator in our class ($\cB_4$) has $L=4$. This means that such a configuration involves at least 16 elementary fields.}.
% Thus $L_2 + L_4 = L_3 + L_1 + 4 \geq 10$. Assuming exactly 10, all pairs $(L_1,L_2) = (3,7),(4,6),(5,5),(6,4),(7,3)$ should yield correlators with non-vanishing contributions of this type, but the number of elementary fields is minimally 16, so there are 8 Wick contractions or more.

A slightly more complicated set-up given in panel (a) of Figure~\ref{fig:topologies}. We have hexagons $123$, $134$, $142$, $243$; only the last does not involve any excitation. We first partition the excitations among $123\sim\alpha$ and $\bar{\alpha}\sim 134\cup 142$, and distribute them among the latter two hexagons, $\bar{\alpha}=\beta\cup\bar{\beta}$. This double-partition yields 
\begin{equation}
\widehat \cA^{(3)} = \sum_{{\bf u} = \alpha \cup \bar \alpha\phantom{\bar \beta}} \sum_{\bar \alpha = \beta \cup \bar \beta} \omega(\alpha,\bar\alpha,l_{13}) \, \omega(\beta,\bar \beta,l_{14}) \ \mathfrak{h}_{123}(\alpha) \, a_{23}^{|\alpha|} \ \mathfrak{h}_{134}(\beta) \, a_{34}^{|\beta|} \ \mathfrak{h}_{142}(\bar \beta) \, a_{42}^{|\bar\beta|} \, .
\end{equation}
Figure~\ref{fig:topologies}~(b) gives a similar amplitude, as it is related to panel~(a) by swapping $2\leftrightarrow4$.

Finally, figure 1~(c) is partitioned in hexagons 143, 132, 123, 134, all of which contain excitations. 
By nesting three partitions we find
\begin{eqnarray}
\widehat \cA^{(4)} = \sum_{\phantom{\bar \beta} {\bf u} = \alpha \cup \bar \alpha} \, \sum_{\bar \alpha = \beta \cup \bar \beta}  \, \sum_{\bar \beta = \gamma \cup \bar \gamma}&&\!\!\!\!\!\!\!
\omega(\alpha,\bar\alpha,l_{13}^{\text{b}}) \, \omega(\beta,\bar \beta,l_{12}) \, \omega(\gamma,\bar \gamma,l_{13}^{\text{f}}) \\  &&  \cdot\,\mathfrak{h}_{143}(\alpha) \, a_{43}^{|\alpha|} \ \mathfrak{h}_{132}(\beta) \, a_{32}^{|\beta|} \ \mathfrak{h}_{123}(\gamma) \, a_{23}^{|\gamma|} \ \mathfrak{h}_{134}(\bar \gamma) \, (a_{34})^{|\bar \gamma|}, \nonumber
\end{eqnarray}
where we indicated by $l_{13}^{\text{b}},l_{13}^{\text{f}}$ the bridge-length between 1 and 3 on the back and front of the ``cushion''.

We can now state our conjecture for the three-level four-point functions with a non-protected operator at $a_1=0$,
\begin{equation}
\langle \cB_{L_1}\cO_{L_2}\cO_{L_3}\cO_{L_4}\rangle=
\sqrt{\frac{L_1 L_2 L_3 L_4}{\mathcal{G} \, S_{21}}}
\left[\sum_{\text{graphs of type }k} \widehat{\mathcal{A}}^{(k)}\right]\,.
\end{equation}
As we remarked, to match our Table~2 we need not to include contributions of the type $\widehat{\cA}^{(2)}$, as they only come in at higher number of Wick contractions. Graphs of type (a) and (b) both contribute to $\widehat{\cA}^{(3)}$, and should be counted separately, cf.~Appendix~\ref{appendix}. Note that, for a fixed set of bridge lengths $l_{jk}$, the amplitudes $\widehat \cA^{(3)}$ and $\widehat \cA^{(4)}$ divided by the root of the $S$ matrix are separately real. In fact, symmetry considerations would in principle still allow us to write arbitrary combinatorial coefficients~$ c^{(3)}(\{l_{jk}\})$ and $ c^{(4)}(\{l_{jk}\})$ in front of each given term. Our proposal amounts to $ c^{(3)}(\{l_{ij}\})=c^{(4)}(\{l_{ij}\})=1$.
This perfectly reproduces the amplitude and space-time dependence of all four-point functions listed in Table~2. In Appendix~\ref{appendix} we explicitly work out two examples for the reader's convenience.

\section{Outlook}

In this letter we have presented a conjecture for computing tree-level four-point functions of single-trace operators in $\cN = 4$ SYM in a special kinematics by integrability methods. We do not use spin-chain scalar products as in Ref.~\cite{Escobedo:2010xs}, but rather the more recent hexagon formalism~\cite{Basso:2015zoa}. The idea of moving the space-time dependence back into the hexagon operator is rather easy to realise for correlators with a single two-excitation BMN operator. We find complete agreement with a number of examples.

There are several natural steps that one might take building on our proposal. Firstly, it would be interesting to consider other non-protected operators, such as twist operators and operators in more general sectors. It would be interesting to see how the symmetry of the Plefka-Drukker vacuum configuration---which, like in the case of three-points, is a diagonal $su(2|2)$ in Beisert's centrally extended $su(2|2)^2$~\cite{Beisert:2005tm}---constrains the four-point function. It would also be  interesting to study the case of more than one non-protected operator, possibly by working out the properties of this kinematics under a suitable ``crossing'' transformation.

Secondly, it would be extremely interesting to explore this kinematics beyond tree-level. 
In field theory, the one-loop correction to the correlators studied in this letter can rather straightforwardly be obtained using two identities~\cite{Eden:1999kh} between the derivatives of certain Feynman integrals. A few one-loop computations of four-point functions involving the Konishi operator are known in the literature~\cite{Bianchi:2001cm,Bianchi:2002rw}; these involve the \textit{singlet} Konishi operator, which cannot be incorporated into the standard integrability scenario. Still, it is interesting to note that the entire one-loop correlator can be expressed by the off-shell four-point one-loop box (the Bloch-Wigner dilogarithm) times some polynomial in cross ratios, and of the very same off-shell box integral with two legs identified. The former part drops in the Drukker-Plefka kinematics, while the latter is logarithmically divergent and contains the information about the anomalous dimensions.  One may ask whether there is a way to generate the anomalous dimension piece from the hexagons; this question concerns the better understood case of three-point functions, too. It is worth noting that in hexagon approach loop corrections generally require accounting for finite-size effects due to \textit{virtual magnons}. One might imagine that precisely these effects reproduce the Bloch-Wigner dilogarithm and the logarithmic divergence expected from field theory.

Furthermore, again from the point of view of field theory, it is worth noting that the structure of the mixed four-point functions is very similar to that pure half-BPS correlators. That case is well-studied, and a classification of integrands on grounds of symmetry and conformal weights has been very successful \cite{Eden:2011we,Eden:2012tu}. For the mixed correlators it should be possible to run the same scheme, though allowing for pseudo-conformal scalar graphs, i.e.\ cases that are divergent due to point identifications. 

Finally, it is intriguing to consider a different set-up where all four operators are half-BPS, but one is displaced by $\epsilon>0$ from the line. Taylor-expanding in $\epsilon$ amounts to populating  with twist operators the spin-chain representing the displaced operator. Were it possible to consider arbitrary-twist operators, we may hope to move away to from the Drukker-Plefka kinematics and describe generic four-point functions. We hope to return to some of these questions in the near future.

\subsection*{Acknowledgements}
We thank Y.~Jiang for interesting related discussions. BE thanks the Institute for Theoretical Physics of ETH Zurich for hospitality during the early stages of this work. AS thanks the organisers of the workshop ``Three-point functions in Gauge/String theories'' at ICTP-SAIFR in  S\~ao Paulo for their hospitality, and the workshop participants for interesting discussions.  BE is supported by the DFG, ``eigene Stelle'' Ed~78/4-3 and acknowledges partial support by the Marie Curie network GATIS under REA Grant Agreement No 317089.  AS's research was partially supported by the NCCR SwissMAP, funded by the Swiss National Science Foundation.

\appendix
\section{Some explicit examples}
\label{appendix}
For the cases of Table~2 we can break down $\sum_{\text{graphs}} \widehat{\mathcal{A}}^{(k)}$ into three parts. For graphs of type (a) we have $\sum_{l_{13},l_{14}} \,\widehat \cA^{(3)}$, for graphs of type (b) we have the same $\sum_{l_{13},l_{14}} \,\widehat \cA^{(3)}$ and for graphs of type (c) we have $\sum_{l_{12},l_{13}^{\text{f}},l_{13}^{\text{b}}} \widehat \cA^{(4)}$. Graphs of type (d) do not appear. Consider for example the simplest case in Table~2, i.e. $G(4;2,2,2)$. Denoting a line by $(ij)$, the three graphs
\begin{equation}
(12) (13) (14)^2 (23), \quad (12) (13)^2 (14) (24), \quad (12)^2 (13) (14) (34),
\end{equation}
are allowed by conformal weight. They can all be drawn on a tetrahedron and in fact are cyclic rotations of each other; they all contribute to amplitude $\widehat{\cA}^{(3)}$. Summing over the associated pairs of edge widths
$(l_{13},l_{14}) = \{(1,2),(2,1),(1,1)\}$ we find
\begin{equation}
G(4;2,2,2) = 2 \left( \frac{4 \sqrt{2} \, a_{34} a_{42}}{\sqrt{3}} + \frac{4 \sqrt{2} \, a_{23} a_{34}}{\sqrt{3}} +\frac{4 \sqrt{2} \, a_{42} a_{23}}{\sqrt{3}} \right) . \label{ex1}
\end{equation}
We highlighted an overall factor of $2$ which is due to the fact that the graphs have a ``handedness'' and therefore can be drawn both on Figure~\ref{fig:topologies}~(a) and on Figure~\ref{fig:topologies}~(b).

For the full set of correlator in Table 2 we proceed as follows: for the topologies of Figure~\ref{fig:topologies}~(a) and Figure~\ref{fig:topologies}~(c), we list all products of propagators that are allowed by conformal weight. Here we insist on topology (c) graphs having $l_{13}^1, l_{13}^2 \geq 1$ so that they cannot be drawn on topology (a). Those graphs of topology (a) that have chirality (i.e.~a square with at least one diagonal) should be counted twice, because they exist and are different on topology (b). However, an empty square or a graph like $(41)^2 (13)^2 (32)^2$ (which is a subgraph of the empty square) does not have chirality and should therefore be put only onto topology (a).% To take it into account a second time as arising from (b) is apparently over-counting. In our examples it was possible to put $c_4 = 1$ for all graphs of topology $(c)$. With this understanding every contributing graph has coefficient $c=1$, though on (a) this has to be doubled for chiral graphs to take into account (b). In general, we cannot solve for all the coefficients as in (\ref{ex1}) because there are only three conditions in equating with Table 2. Yet, our rule $c=1$ is perfectly natural.

Let us work out a more slightly more complicated example.
Consider $G(5;2,5,2)$. For topology (a) we have the candidates
\begin{eqnarray}
&& c_{3,1} \, (13)^5 (24)^2, \nonumber \\
&& c_{3,2} \, (13)^3 (14)^2 (23)^2, \ c_{3,3}\, (12)^2 (13)^3 (34)^2, \nonumber \\
&& c_{3,4} \, (13)^4 (14) (23) (24), \ c_{3,5} \, (12) (13)^4 (24) (34), \\
&& c_{3,6} \, (12) (13)^3 (14) (23) (34) \, ,\nonumber
\end{eqnarray}
where we have allowed arbitrary coefficients for all graphs for the time being.
Here the first graph is disconnected. Upon evaluation it vanishes as it should because $\cB_5(0)$ and $\cO_5(a_3)$ carry different $SU(4)$ representations. Graphs 2,3 are of ``sausage type'' whereas 4,5 are empty squares. In fact, the $\hat \cA^{(3)}$ amplitudes for these two empty squares also vanish. Thinking about colour factors this seems reasonable, because the colour structure of the lines (14)(42)(23) or (12)(24)(43), respectively, going through the two half-BPS operators is simply $\delta^{ab}$ and the BPS/BPS propagator in the middle is equal to 1. The structure of the graphs is therefore very similar to the disconnected case. Upon evaluation we thus find from topology (a):
\begin{equation}
G(5;2,5,2)_{(a)} = c_{3,2} \, \sqrt{10} \, a_{34}^2 + c_{3,3} \, \sqrt{10} \, a_{23}^2  - c_{3,6} \sqrt{10} \, a_{23} a_{34}.
\end{equation}
In this case we also have type (c) graphs. In general we will input all six permutations of the operators at points 2,3,4 and divide by 2 in order to compensate for the rotation symmetry of the topology around the $13$ axis. Here one can only construct candidates for $\cO_5$ at point 3, though. There are six cases:
\begin{eqnarray}
&& c_{4,1} \, (14)^2 (23)^2 \bigl((13)_{\text{b}}\bigr)^2 (13)_{\text{f}}, \ c_{4,2} \, (12)^2 (34)^2 \bigl((13)_{\text{b}}\bigr)^2 (13)_{\text{f}}, \nonumber \\
&& c_{4,3} \, (14)^2 (23)^2 (13)_{\text{b}} \bigl((13)_{\text{f}}\bigr)^2, \ c_{4,4} \, (12)^2 (34)^2 (13)_{\text{b}} \bigl((13)_{\text{f}}\bigr)^2, \\
&& c_{4,5} \, (12) (14) (23) (34) \bigl((13)_{\text{b}}\bigr)^2 (13)_2, \ c_{4,6} \, (12) (14) (23) (34) (13)_{\text{b}} \bigl((13)_{\text{f}}\bigr)^2 .\nonumber
\end{eqnarray} 
The graphs 1,2,3,4 are of equal topology and should have equal coefficients. It turns out that all of them give the same result even though points 1,3 are not equivalent. The factor of 1/2 for flipping points 2,4 has already been taken into account. Graphs 5,6 also give equal contributions due to this symmetry. We find
\begin{equation}
G(5;2,5,2)_{(c)} = (c_{4,2} + c_{4,2} + c_{43} + c_{4,4}) \sqrt{\frac{5}{2}} \, \bigl( a_{23}^2 + a_{34}^2 \bigr) + (c_{4,5} + c_{4,6}) \sqrt{10} \, a_{23} a_{34}.
\end{equation}
On grounds of symmetry we could only assert $c_{4,i} = c_4 : i \in \{1,2,3,4\}, \, c_{4,j} = \hat c_4 : i \in \{5,6\}$ leaving 5 parameters. Our prescription amounts to imposing $c_{3,2} = c_{3,3} = c_{4,i} = 1,$ and $c_{3,6} = 2$, where the last condition accounts for the fact that the graph  with coefficient $c_{3,6}$ contributes to both type (a) and (b). Then
\begin{equation}
G(5;2,5,2) = 3 \sqrt{10} \, \bigl(a_{23}^2 + a_{34}^2\bigr),
\end{equation}
in full agreement with Table 2.

\bibliography{refs}{}
\bibliographystyle{nb}

\end{document}